\documentstyle[fleqn]{article}
\begin{document}
\newcommand{\su}{\mbox{$\sf U$}} 
\newcommand{\sm}{\mbox{$\sf M$}}
\newcommand{\sn}{\mbox{$\sf N$}}
\newcommand{\snpri}{\mbox{${\sf N}^\prime$}}
\newcommand{\ssf}{\mbox{$\sf F$}} 
\renewcommand{\sl}{\mbox{$\sf L$}}
\newcommand{\soneN}{\mbox{${\sf I}_N$}}
\newcommand{\soned}{\mbox{${\sf I}_2$}}   
\newcommand{\sonef}{\mbox{${\sf I}_4$}}   
\newcommand{\shk}{\mbox{${\sf H}_k$}}
\newcommand{\spk}{\mbox{${\sf P}_k$}}
\newcommand{\spp}{\mbox{${\sf P}_+$}}
\newcommand{\spm}{\mbox{${\sf P}_-$}}
\newcommand{\sppm}{\mbox{${\sf P}_\pm$}}
\newcommand{\va}{\mbox{$\vec A$}}
\newcommand{\vb}{\mbox{$\vec B$}}
\newcommand{\vc}{\mbox{$\vec C$}}
\newcommand{\vd}{\mbox{$\vec D$}}
\newcommand{\vl}{\mbox{$\vec L$}}
\newcommand{\vm}{\mbox{$\vec M$}}
\newcommand{\vmu}{\mbox{$\vec \mu$}}
\newcommand{\vn}{\mbox{$\vec N$}}
\newcommand{\vnpri}{\mbox{$\vec {N^\prime}$}} 
\newcommand{\vnu}{\mbox{$\vec \nu$}}
\newcommand{\vecf}{\mbox{$\vec f$}}
\newcommand{\vecr}{\mbox{$\vec R$}} 
\newcommand{\vrho}{\mbox{$\vec \rho$}} 
\newcommand{\vsl}{\mbox{$\vec {\sf \Lambda}$}}
\newcommand{\slambdaj}{\mbox{${\sf \Lambda}_j$}}
\newcommand{\slambdak}{\mbox{${\sf \Lambda}_k$}}
\newcommand{\slambdal}{\mbox{${\sf \Lambda}_l$}}
\newcommand{\slambdan}{\mbox{${\sf \Lambda}_n$}} 
\newcommand{\eA}{\mbox{${\vec e}_\alpha$}}
\newcommand{\eB}{\mbox{${\vec e}_\beta$}}
\newcommand{\eG}{\mbox{${\vec e}_\gamma$}}
\newcommand{\vA}{\mbox{$\vec \alpha$}}
\newcommand{\vB}{\mbox{$\vec \beta$}}
\newcommand{\vG}{\mbox{$\vec \gamma$}}
\newcommand{\vS}{\mbox{$\vec {\sf \sigma}$}}
\newcommand{\ox}{\mbox{$\otimes$}}
\newcommand{\od}{\mbox{$\odot$}}
\newcommand{\be}{\begin{equation}}
\newcommand{\bea}{\begin{eqnarray}} 
\newcommand{\ee}{\end{equation}} 
\newcommand{\eea}{\end{eqnarray}} 
\title{Baker-Campbell-Hausdorff relation for special unitary groups $SU(N)$}
\author{Stefan Weigert\cite{thanks}\\
	Departement f\"ur Physik und Astronomie der Universit\"at Basel\\
	Klingelbergstrasse 82, CH-4056 Basel}
\date{October 7, 1997}
\maketitle
\begin{abstract}
Multiplication of two elements of the special unitary group $SU(N)$ 
determines uniquely a third group element. A Baker-Campbell-Hausdorff 
relation is derived which expresses the group parameters of the product 
(written as an exponential) in terms of the parameters of the exponential 
factors. This requires the eigenvalues of three $(N\times N)$ matrices. 
Consequently, the relation can be stated analytically up to $N=4$, in 
principle. Similarity transformations encoding the time evolution of quantum 
mechanical observables, for example, can be worked out by the same means.
\end{abstract} 
\section{Introduction}
Various questions in physics reduce to the following problem: write the 
product of exponential functions depending on noncommuting operators 
$\widehat A$ and $\widehat B$, respectively, as the exponential of a third 
operator, $\widehat C$, 
\be
\exp [ {\widehat A} ] \exp [ {\widehat B} ] 
	     = \exp [ {\widehat C} ] \, .
\label{compose}
\ee
The names of Baker, Campbell, and Hausdorff (BCH) are associated (cf.\ 
\cite{wilcox67}) with a formula for the operator $\widehat C $ expressed in 
multiple commutators of $\widehat A$ and $\widehat B$:
\be
{\widehat C} = \widehat A + \widehat B 
	       + \frac{1}{2} [\widehat A , \widehat B ] 
	       + \frac{1}{12} \left([\widehat A , [ \widehat A , \widehat B]] 
		      + [[ \widehat A , \widehat B], \widehat B ] \right) 
	       + \ldots
\label{defC}
\ee
Remarkably, the operator $\widehat C$ is expressed depends on commutators 
of $\widehat A$ and $\widehat B$ {\em only} implying that it is contained in 
the same algebra as $\widehat A$ and $\widehat B$. For this result to hold 
it is crucial to consider products of {\em exponential} functions. 

Although the expansion (\ref{defC}) for the operator $\widehat C$ is 
explicit, usually the infinite series of repeated commutators cannot be 
summed in closed form. It may be used, however, to generate an approximate  
expression for $\widehat C$ by directly calculating a finite number of terms
\cite{erikson68}. 
When read from left to right, Eq.\ (\ref{compose}) shows how to 
{\em entangle} the two factors into a single exponential. An application
important in quantum mechanics results for the Heisenberg group of position 
and momentum operators $\hat q$ and $\hat p$, where
\be
\exp [ -i {\hat p}] \exp [ -i {\hat q} ]
	     = \exp [ -i ( {\hat p} + {\hat q}) + i\hbar/2 ] \, .
\label{HWBCH}
\ee
The right-hand-side is particularly simple because the commutator
\be
[ {\hat p} , {\hat q} ] = \frac{ \hbar }{i} 
\label{pq}
\ee
is a constant such that only the first commutator in (\ref{defC})
contributes to the operator ${\widehat C}$. Another situation with 
the need for entangling two operators is encountered in periodically driven 
systems. In specific cases, the propagator over one full period 
reduces to a product of the propagators for shorter intervals 
\cite{berry+79,casati+79,schatzer+97}. The Lie algebras involved in these  
`quantum maps' may either have a finite or an infinite number of elements. 
 
When read in the opposite sense, Eq. ({\ref{compose}) represents a 
{\em disentangling} relation, that is, the decomposition of a single 
exponential into factors with simple properties. Such a relation is useful to 
calculate expectation values of basic operators in the group $SU(2)$, for 
example, since they are easily derived from a generating function in 
disentangled form \cite{arecchi+72}. Similarly, changes of the group 
parameterization \cite{gilmore74a} are conveniently performed by using 
BCH-relations. In general, the discussion of coherent states for 
particle and spin systems as well as for arbitrary Lie groups 
\cite{perelomov86} benefits from the knowledge of (de-)composition rules 
(\ref{compose}). 

A closely related question arises from the need to perform similarity 
transformations according to 
\be
\exp [ - {\widehat A} ]  {\widehat B}  \exp [ {\widehat A} ] 
		 = {\widehat B}^\prime \, .
\label{sim}
\ee
If the operator $\widehat A$ is proportional to $i$ times the Hamiltonian of 
a quantum system, Eq. (\ref{sim}) describes the time evolution of the 
Heisenberg observable $\widehat B$ into ${\widehat B}^\prime$.

A number of techniques has been established in order to efficiently treat  
entangling and disentangling problems, in particular, if the operators 
involved in the BCH-relation are elements of a {\em finite}-dimensional Lie  
algebra. Two-dimensional unitary faithful irreducible representations are 
used to derive explicit results for the group $SU(2)$ \cite{gilmore74a}, and for the group of the 
harmonic oscillator \cite{miller72,gilmore74b}, for example. Applications to 
more complicated cases involving symplectic groups also have been worked out 
in detail \cite{twamley93,gilmore+89}. However, it is {\em not} necessary to 
exclusively work with unitary representations: any faithful representation 
can be used \cite{gilmore74b}. This is helpful if one knows a representation 
consisting of upper and lower triangular matrices since they are easily 
exponentiated. Disentanglement of Lie group elements is also achieved by  
using recursion relations for expanded exponentials and Laplace-transform 
techniques \cite{rhagunathan+92}. This approach generalizes a method first 
applied to particular group elements of $SU(3)$ \cite{rhagunathan+89}.
The powerful approach in \cite{wilcox67} maps the problem of both 
(dis-)entangling (\ref{compose}) and similarity (\ref{sim}) transformations 
to the solution of a set of coupled first-order differential equations. This 
paper also contains theoretical background on BCH-relations, applications in 
physics as well as a large number of references.

In the present paper a different method to evaluate BCH-relations is 
developped for the groups $SU(N)$. It is based on the spectral theorem for 
hermitian operators in finite-dimensional vector spaces. A `linearized' 
version 
of this result is derived by exploiting a specific feature of the algebra 
$su(N)$ going beyond its Lie algebraic properties. In this way, a one-to-one 
correspondence between an exponential of linearly combined generators and a 
linear combination of them is established -- `removing' thus the exponential 
function. It is then straightforward to entangle elements of the group 
$SU(N)$. Conceptually, this method is related to work performed in the early 
1970's where the study of chiral algebras required the evaluation of {\em 
finite} transformations for special unitary groups 
\cite{barnes+72a,barnes+72b}. In that context, however, BCH-relations have 
not been considered.

\section{Some Fundamentals of $SU(N)$}

An irreducible faithful representation of the group $SU(N)$ \cite{itzykson+66} 
is given by the set of all unitary $(N \times N)$ matrices $\sf U$ with unit 
determinant, 
\be
\det \su = 1, \qquad \su_{nn'} \in {\bf C}  \, , 
	      \quad n,n' = 1, \ldots, N.
\label{det}
\ee
Each matrix \su\ can be written in the form
\be
\su = \exp [ -i \sl ]\, , \qquad \sl^\dagger = \sl\, ,
\label{uform}
\ee
with a traceless hermitian matrix \sl . It is conveniently expressed as a 
linear combination 
\be 
\sl = \vl \cdot \vsl 
	   \equiv \sum_{j=1}^{N^2 -1} L_j \slambdaj \, ,
	  \qquad  L_j \in {\bf R} \, ,
\label{lvector}
\ee
with the set  \vsl\ forming a basis for traceless hermitian matrices,
$\slambdaj^\dagger = \slambdaj$, called the generators. At the same time, 
they are a basis of the Lie {\em algebra} $su(N)$ of $SU(N)$, satisfying the 
commutation relations:
\be
\left[ \slambdaj , \slambdak \right]_- 
		      = 2 i f_{jkl} \slambdal \, , 
\label{comm}                      
\ee
where the indices $j,k,l,$ take values from $1$ to $N^2 -1$, the summation 
convention for repeated indices applies, and the $(N\times N)$ unit matrix is 
denoted by $\soneN$. The structure constants $f_{jkl}$ are elements of a 
completely antisymmetric tensor (spelled out explicitly in \cite{greiner+90}, 
for example) with Jacobi identity   
\be
f_{klm} f_{mpq} + f_{plm} f_{mkq} + f_{kpm} f_{mlq} = 0 \, . 
\label{jaco1} 
\ee
The group $SU(N)$ has rank $(N-1)$. In other words, any maximal abelian 
subalgebra of $su(N)$ consists of $(N-1)$ elements corresponding to 
all linearly independent traceless $N$-dimensional diagonal 
matrices. A `complete set of commuting variables' for a quantum system 
described by $SU(N)$ would contain in addition the same number of Casimir 
operators according to Racah's theorem \cite{greiner+90}.

A particular feature of the algebra $su(N)$ is its closure under 
{\em anti\ }commutation of its elements:
\be
\left[ \slambdaj , \slambdak \right]_+  
		      = \frac{4}{N} \delta_{jk} \, \soneN 
			   + 2 d_{jkl} \slambdal \, , 
\label{anticomm}
\ee
where the $d_{jkl}$ form a totally symmetric tensor (cf.\ 
\cite{greiner+90}). For $N=2$, {\em all} numbers $d_{jkl}$ are equal to 
zero, and the generators  \vsl\ coincide with the Pauli matrices \vS : the 
anticommutator of two of them is either equal to zero or a multiple of the 
unit matrix, $\soned$. 

The anticommutation relation will be important in the present context but it 
is {\em not} generic for an arbitrary Lie algebra. As a consequence of 
(\ref{anticomm}), two generators \slambdaj\  and  \slambdak\  
of $su(N)$ are `orthogonal' to each other with respect to the trace: 
\be
\mbox{Tr} (\slambdaj  \slambdak ) = 2 \delta_{jk} \, .
\label{orthon}
\ee
In addition, a second Jacobi-type identity exists involving both the 
antisymmetric and the symmetric structure coefficients in (\ref{comm}) 
and (\ref{anticomm}): 
\be
f_{klm} d_{mpq} + f_{kqm} d_{mpl} + f_{kpm} d_{mlq} = 0 \, .
\label{jaco2}
\ee

For the following, a vector-type notation is useful, based on the structure 
constants and the Kronecker symbol. Define the scalar product as employed 
already in Eq.\ (\ref{lvector}), 
\be
\va \cdot \vb = A_n \delta_{nm} B_m = A_n B_n \, , 
\label{spro}
\ee
where the components of \va\ and  \vb\  are allowed to be either numbers or 
generators \slambdan . Similarly, define an antisymmetric `cross product' 
$\ox$ by
\be
(\va \ox \vb )_j = f_{jkl} A_k B_l = - (\vb \ox \va)_j \, ,
\label{cross}
\ee
and a symmetric `dot product' $\od$:
\be
(\va \od \vb )_j = d_{jkl} A_k B_l = + (\vb \od \va)_j \, .
\label{dot}
\ee
Then, the relations (\ref{comm},\ref{anticomm}) can be written
\bea
\left[ \va \cdot \vsl , \vb \cdot \vsl \right]_- 
		  &=& 2i (\va \ox \vb) \cdot \vsl \, , \\
\left[ \va \cdot \vsl , \vb \cdot \vsl \right]_+ 
		  &=& \frac{4}{N} \va \cdot \vb \, \soneN 
		  + 2 ( \va \od \vb) \cdot \vsl \, , 
\label{comm2}
\eea
where \va\ and  \vb\  are {\em arbitrary} vectors of dimension $(N^2 -1)$ 
with numeric entries. Adding these equations leads to a compact form of the   
(anti-) commutation relations:
\be
(\va \cdot \vsl)  (\vb \cdot \vsl) 
	     = \frac{2}{N} \va \cdot \vb \, \soneN 
	       + ( \va \od \vb +i \va \ox \vb ) \cdot \vsl \, .
\label{comcom}
\ee
This equation emphasizes the important point that any expression 
{\em quadratic} in the generators can be expressed as a {\em linear} 
combination of them, including the identity. As a matter of fact, it 
generalizes the known identity in $SU(2)$ for the Pauli matrices:
\be
(\va \cdot \vS)  (\vb \cdot \vS) 
	     = \va \cdot \vb \, \soneN + i \va \ox \vb  \cdot \vS \, .
\label{paulicom}
\ee

In the new notation, the identities (\ref{jaco1},\ref{jaco2}) read  
\bea
  (\va \ox \vb ) \cdot ( \vc \ox \vd ) 
+ (\vc \ox \vb ) \cdot ( \va \ox \vd ) 
+ (\va \ox \vc ) \cdot ( \vb \ox \vd ) &=& 0 \, , \label{newjaco1}   \\
  (\va \ox \vb ) \cdot ( \vc \od \vd ) 
+ (\va \ox \vd ) \cdot ( \vc \od \vb ) 
+ (\va \ox \vc ) \cdot ( \vb \od \vd ) &=& 0 \, .  
\label{newjaco2}                               
\eea
Another useful form of Eq. (\ref{jaco2}) is given by
\be
  \va \ox (\vb \od \vc ) 
=  (\va \ox \vb ) \od \vc + \vb \od (\va \ox \vc ) \, ,  
\label{deriv}
\ee
showing that applying $\va \ox $ to a $\od$ product acts as does a derivative. 
The `orthogonality' of the generators (\ref{orthon}) becomes
\be
\mbox{Tr} \left( ( \va \cdot \vsl ) ( \vb \cdot \vsl ) \right) 
= 2 \va \cdot \vb \, ,   
\label{ortho2}
\ee
for arbitrary \va\ and \vb .

\section{Spectral theorem}

Every matrix $\sm \in {\bf C}^N$ satisfies its own characteristic 
equation,
\be
\sum_{n=0}^{N} a_n \sm^n = 0 \qquad a_{N}=1 , \, a_0 = \det \sm \, , 
\label{char}
\ee
according to the theorem of Cayley-Hamilton. The coefficients $a_n$
define the characteristic polynomial of \sm . For traceless matrices 
such as \sm\ $\in su(N)$, the coefficient $a_{N-1}$ in Eq.\ (\ref{char}) 
is equal to zero since it equals the trace of \sm . 
According to  Eq.\ (\ref{char}), any power $N'\geq N$ of the 
matrix \sm\ is identical to a linear combination of its powers 
$\sm^n$ with $0 \leq n 
\leq N-1$. The expansion of a matrix exponential can thus be written  
\be
\exp [ -i \sm ] = \sum_{m=0}^\infty \frac{(-i \sm)^m}{m!} 
		= \sum_{n=0}^N e_n(\sm) \sm^n \, ,
\label{expo}
\ee
with uniquely defined coefficients $e_n(\sm)$. They are determined 
directly by referring to the {\em spectral theorem} \cite{halmos58} valid for 
smooth functions $f$ of a hermitian matrix \sm\ with (nondegenerate) 
eigenvalues $m_k$, $k=1,\ldots,N$:
\be
f(\sm ) = \sum_{k=1}^N f(m_k) \spk \, ,
\label{spectr}
\ee
and the operator $\spk = | m_k\rangle \langle m_k|$ projects down to the 
one-di\-men\-si\-onal eigen\-space spanned by the eigenvector $|m_k\rangle$ 
associated with the eigenvalue $m_k$. In terms of powers $\sm^k$ and the 
eigenvalues $m_k$, the matrices \spk\ read     
\be
\spk = \prod_{n\neq k} \frac{\sm -m_n}{m_k-m_n} 
     = \sum_{n=0}^{N-1} P_{kn} \sm^n \, ;
\label{project}
\ee
the sum contains powers $\sm^{N-1}$ at most since the product runs over 
$(N-1)$ factors. Combining Eqs. (\ref{spectr}) and (\ref{project}), one 
obtains that
\be
f( \sm ) =       \sum_{n=0}^{N-1} 
		     \left( \sum_{k=1}^N P_{kn} f(m_k) \right) \sm^n 
	 \equiv  \sum_{n=0}^{N-1} f_n \sm^n \, ,
\label{coeff}
\ee
and, upon choosing $f(x) \equiv \exp (-ix)$, the sum in the round brackets 
produces the coefficients $e_n$ of the expansion (\ref{expo}) in terms of the 
eigenvalues $m_k$. 

It is possible \cite{rhagunathan+92} to express the numbers $f_n$ in 
(\ref{coeff}) differently. Write the coefficient $f_{N-1}(\sm, \lambda)$ of 
$\sm^{N-1}$ with a dummy parameter $\lambda$ introduced as follows:
\be
f_{N-1}(\sm , \lambda) = \sum_{n=1}^N \Delta_n f(\lambda m_k) \, , 
    \qquad \Delta_n = \prod_{k\neq n}(m_n - m_k)^{-1} \, .
\label{highk}
\ee
Linear combinations of derivatives with respect to $\lambda$ yield
the remaining coefficients $f_n$, $n=0,1,\ldots, N-2$, associated with
any smooth function $f$:
\be
f_n(\sm) = \left[ \left( \partial_\lambda^{N-n-1} 
		  - \sum_{\nu=1}^{N-n-1} a_{N-\nu} 
			  \partial_\lambda^{N-n-1-\nu} \right) 
			  f_{N-1}(\sm,\lambda) \right]_{\lambda=1} \, , 
\label{lowk}
\ee
with numbers $a_n$ from the characteristic polynomial (\ref{char}), and the 
abbreviaton $d/d\lambda \equiv \partial_\lambda$. Since  Eq.\ (\ref{coeff}) 
requires the eigenvalues of \sm, analytic expressions will be obtained only 
for $(4\times4)$ matrices at most, i.e. for $SU(4)$.

\section{Linearized spectral theorem}

A stronger version of relation (\ref{spectr}) is derived now. It is valid for  
for hermitian $(N \times N)$ matrices, and it will be called the 
{\em linearized spectral theorem}:
\be
f( \vm \cdot \vsl ) = f_0 ( \vm ) \soneN + \vecf (\vm ) \cdot \vsl \, .
\label{lst}
\ee
It states that any function $f$ of a linear combination of the 
generators  \vsl\ of $SU(N)$ is equal to a linear combination of the 
identity and the generators with well-defined coefficients $(f_0,\vecf)$.
In other words, the powers of the generators  \vsl\ contained in the powers 
$\sm^n\equiv (\vm \cdot \vsl)^n$ in  Eq.\ (\ref{coeff}) can be reduced to linear combinations of them. 
In view of the commutation relations of the algebra $su(N)$,  Eq.\ 
(\ref{comcom}), this is not surprising: the required reduction is carried 
out in a finite number of steps by repeatedly expressing products of two 
generators by a linear combination of generators. 

A convenient procedure to determine $(f_0,\vecf )$ in (\ref{lst}) starts 
from writing
\be
\sm^n = \mu_{0,n} \, \soneN + \vmu_n \cdot \vsl \, ,  
      \qquad n=0,1,2 \ldots, N-1 \, , 
\label{Mpower}
\ee
where
\bea
\mu_{0,0} &=& 1 \, , \quad \mu_{0,1} = 0 \, ,
			       \label{start0} \\
\vmu_0 &=& 0 \, , \quad \vmu _1 = \vm \, .   
\label{start}
\eea
A recursion relation for $( \mu_{0,n}, \vmu_n)$ follows from writing
$\sm^{n+1} = \sm^n \sm$, using   (\ref{comcom}) and (\ref{Mpower}),
\bea
\sm^{n+1} &=& \mu_{0,n} \vm \cdot \vsl + ( \vmu_n \cdot \vsl )(\vm \cdot\vsl) 
						       \nonumber \\
	  &=& \frac{2}{N} \vmu_n \cdot \vm \, \soneN + 
	     (\mu_{0,n} \vm + \vmu_n \od \vm + i \vmu_n \ox \vm ) \cdot \vsl \, .
\label{recurse}
\eea
Comparison with (\ref{Mpower}) for $(n+1)$ instead of $n$ shows that
\bea
\mu_{0,n+1} &=& \frac{2}{N} \vmu_n \cdot \vm \, , 
\label{red0} \\
\vmu_{n+1}  &=& \mu_{0,n} \vm + \vmu_n \od \vm + i \vmu_n \ox \vm 
	    = \frac{2}{N} (\vmu_{n-1} \cdot \vm ) \vm + \vmu_n \od \vm\, , 
\label{red1}
\eea
which recursively defines $(\mu_{0,n}, \vmu_n)$ in terms of \vm , starting 
with the `initial values' (\ref{start0},\ref{start}). The terms $i \vmu_n \ox \vm$ 
do {\em not} contribute since each $\vmu_n$ following from (\ref{Mpower}) is 
proportional to $\vm , \vm \od \vm , (\vm \od \vm ) \od \vm, \ldots$ Using 
the derivative-like property (\ref{deriv}), one always encounters terms  
$\vm \ox \vm$ being equal to zero. Consequently, the coefficients 
$(f_0, \vecf)$ on the right-hand-side of (\ref{lst}) have been expressed 
explicitly through  \vm\ and the eigenvalues $m_k$:
\be
f_0 (\vm ) = \sum_{n=0}^{N-1} f_n \mu_{0,n} \, , 
\qquad \vecf ( \vm ) = \sum_{n=0}^{N-1} f_n \vmu_n \, ,
\label{ff}
\ee
with $f_n$ from Eqs. (\ref{highk}) and (\ref{lowk}). 
Note that according to 
(\ref{red1}) the expression for $\vecf (\vm )$ contains only totally 
symmetric powers $\vm , \vm \od \vm$, $ (\vm \od \vm ) \od \vm, \ldots$
Given \vm , a simple expression for $f_0$ is provided by taking the trace 
of  Eq.\ (\ref{lst}):
\be 
f_0 ( \vm ) = \frac{1}{N} \mbox{Tr}( f( \vm \cdot \vsl ) )
	     = \frac{1}{N} \sum_{k=1}^N f (m_k) \, .
\label{f0}
\ee
It should be pointed out that $f_0$ is {\em not} independent of \vecf : 
one can solve the recursion for $\vmu_n$, Eq.\ (\ref{red1}) without refering
to (\ref{red0}). This is reasonable because only then one has the {\em same} 
number of parameters in \vm\ and on the right-hand-side of (\ref{lst}). 

Suppose now that the {\em right-hand-side} of  Eq.\ (\ref{lst}) is given, i.e.
the parameters $(f_0, \vecf)$ are known to define a group element of $SU(N)$. 
How does one express \vm\ in terms of \vecf ? This is actually the difficult 
step when deriving a BCH-formula: to find the group element in terms of the 
the original parametrization. 
Assume the function $f$ to be invertible, then one can write
\be
\vm \cdot \vsl = f^{-1} (f_0 \soneN +\vecf \cdot \vsl) = F ( \vecf \cdot \vsl)
\label{invert}
\ee
with a new function $F$. The clue to the inversion is to realize that 
(\ref{invert}) represents an equation of the type (\ref{lst}) again. This 
follows from reading Eq. (\ref{lst}) from right to left, replacing $f\to F$, 
exchanging the role of \vecf\ and \vm , and setting $f_0$ equal to zero in 
(\ref{lst}). Now the reasoning leading to  Eq.\ (\ref{ff}) can be repeated 
in order to determine $\vm = \vm ( \vecf )$. Therefore, \vm\ can be found as 
a function of \vecf\ by the means already established. 

The orthonormality (\ref{orthon}) for the generators  \vsl\ allows one to 
formally switch from  \vm\ to  \vecf\ and {\em vice versa} in a simple 
manner: multiply  Eq.\ (\ref{lst}) with  \slambdak\ and take the trace which 
leads to 
\be
f_k = \mbox{Tr} \left( f_0 \slambdak + \vecf \cdot \vsl \slambdak \right) 
    = \mbox{Tr} \left( f( \vm \cdot \vsl ) \slambdak \right) \, ,
\label{fk}
\ee
while the inverse transformation follows from (\ref{invert}):
\be
M_k = \mbox{Tr} \left( (\vm \cdot \vsl ) \slambdak \right)
    = \mbox{Tr} \left( f^{-1} ( f_o \, \soneN + \vecf \cdot \vsl) \slambdak 
				      \right) \, .
\label{Mk}
\ee

Before applying the linearized spectral theorem to the derivation of 
BCH-formulae, a comment on the relation between the matrices 
$\sm = \vm \cdot \vsl$ and  \ssf\    in (\ref{lst}),  
\be
f( \sm ) = f_0 \, \soneN + \ssf
\label{lstip}
\ee
should be made. One must have $[ \sm, \ssf ] = 0$ since Eq. (\ref{lstip}) 
is an identity. Nevertheless, the matrices involved do
not have to be multiples of each other. The vanishing commutator implies 
that the matrices \sm\ and \ssf\ can be diagonalized simultaneously. Having 
done this \sm\ would be given by a specific linear combination of $(n-1)$
traceless diagonal generators $\shk , k=1,2, \ldots, N-1$. The matrix  \ssf\  
commutes with \sm\ and it is therefore only required to be another element of 
the maximal abelian subalgebra containing \sm . For the group $SU(2)$, the 
dimension of this algebra is equal to one: \sm\ and  \ssf\  are in this (and 
only this) case proportional to each other (cf. the first example below). 
For $SU(3)$ this observation is illustrated by a result of \cite{bulgac+90} 
where Lie groups are studied from a geometric point of view. In an 
appropriate local basis, any group element can be written as a function of  
a linear combination of two commuting operators which span a maximal abelian   
subalgebra.

\section{BCH for $SU(N)$}

A Baker-Campbell-Hausdorff relation for composing of group elements of 
$SU(N)$ follows from twofold application of the linearized 
spectral theorem with $f(x) = \exp [-i x ]$. Consider 
the product of two finite transformations, $\exp [ -i \vm \cdot \vsl ]$ and 
$\exp [ -i \vn \cdot \vsl ]$, which defines a third element of $SU(N)$ 
characterized by \vecr ,
\be
\exp [ -i \vecr \cdot \vsl ] 
= \exp [ -i \vm \cdot \vsl ] \exp [ -i \vn \cdot \vsl ] \, .
\label{mult}
\ee
 Using  Eq.\ (\ref{lst}) with the exponential function, 
one obtains
\bea
\exp [ -i \vecr \cdot \vsl ]
&=& \mu_0 \nu_0 \soneN 
   + ( \nu_0 \vmu + \mu_0 \vnu ) \cdot \vsl  
   + (\vmu \cdot \vsl ) ( \vnu \cdot \vsl ) \nonumber \\
&=& (\mu_0 \nu_0 + \frac{2}{N} \vmu \cdot \vnu ) \soneN 
  + ( \nu_0 \vmu + \mu_0 \vnu + \vmu \od \vnu + i \vmu \ox \vnu ) \cdot \vsl 
					 \nonumber \\ 
&=& \rho_0 \soneN + \vrho \cdot \vsl \, , 
\label{comp}
\eea
using the commutation relations (\ref{comcom}). The quantities 
$(\rho_0,\vrho)$ can be read off directly as the coefficients of \soneN\ 
and \slambdaj , respectively. The components of  
\vecr\  are thus given by  Eq.\ (\ref{Mk}):
\bea
R_k &=& i \, \mbox{Tr} \left\{ \ln \left[ 
 (\mu_0 \nu_0 + \frac{2}{N} \vmu \cdot \vnu ) \soneN \right. \right.
					      \nonumber \\ 
& &  \left. \left. + ( \nu_0 \vmu + \mu_0 \vnu + \vmu \od \vnu 
					       + i \vmu \ox \vnu ) 
  \cdot \vsl \right] \slambdak \right \} \, , 
\label{Rk}
\eea
providing the relation $\vecr = \vecr ( \vm, \vn ) $. The explicit evaluation 
requires diagonalization of the matrices $ \sm$ an $ \sn$ in order to 
determine \vmu\ and \vnu ; finally, $\vrho \cdot \vsl$ has to be diagonalized 
in order to get rid of the logarithm in  Eq.\ (\ref{Rk}). In total, three 
$(N \times N)$ matrices have to be diagonalized to achieve the entangling.

As an illustration, the familiar example of $SU(2)$ will be looked at from 
the point of view developed here. However, the $\od$ product being identical 
to zero, this case does not exhibit the full complexity. Therefore, $SU(4)$ 
will also be discussed briefly. Before giving the examples, the use of the 
linearized spectral theorem for the determination of similarity 
transformations in the group $SU(N)$ will be indicated.  

\section{Similarity transformations}
The transformation of the operator $\sn = \vn \cdot \vsl \in su(N)$ under 
$\sm= \vm \cdot \vsl \in su(N)$ according to 
\be 
\exp [-i \sm ] \, \sn \, \exp[ i \sm ] = \snpri
\label{simil}
\ee
could be determined from the linearized spectral theorem in the following way. 
Write the group element as
\be
\exp[ i \sm ] = \mu_0 \soneN + \vmu \cdot \vsl \, ,
\label{exponential}
\ee
and its inverse follows from the adjoint of this equation as
\be
\exp[ - i \sm ] = {\mu_0}^* \soneN + {\vmu}^* \cdot \vsl \, ,
\label{inverseexponential}
\ee
where the star denotes complex conjugation. Plugging these expressions into 
(\ref{simil}), one encounters triple products of generators \vsl\ which when 
reduced to a linear combination lead to a somewhat involved expression. It is   
more convenient to first multiply Eq. (\ref{simil}) with $\exp[ i \sm ]$,
and to work out the terms {\em quadratic} in the generators. Comparison of 
the coefficients of \soneN\ and \vsl\ leads to  
\bea
&&\vmu \cdot \vnu = \vmu \cdot \vnpri \, , 
\label{scalar}\\
&&\mu_0 \vn + \vn \od \vmu + i \vn \ox \vmu 
       = \mu_0 \vnpri + \vmu \od \vnpri + i \vmu \ox \vnpri \, .
\label{vector}        
\eea
It is the vector \vnpri\ which must be determined from these equations. It 
is useful to rewrite Eq. (\ref{vector}) with matrices  
\be
{\bf K}_\pm \equiv \mu_0 \soneN + \vmu \od  \pm i \vmu \ox  \, ,
\label{matrixK}
\ee
acting on the vectors \vn\ and \vnpri , respectively,
\be
{\bf K}_- \vn = {\bf K}_+ \vnpri \, .
\label{vectorshort}
\ee
The matrix ${\bf K}_+$ {\em does} have an inverse, ${\bf K}_+^{-1}$, since it 
describes the action of $\exp [ i \sm ]$ on \snpri\ which {\em is} invertible. 
Consequently, the vector \vnpri\ is determined by the relation
\bea 
\vnpri &=& {\bf K}_+^{-1} {\bf K}_- \vnu \nonumber \\
	&=& \left( \mu_0\soneN + \vmu \od  + i \vmu \ox \right)^{-1} 
	    \left( \mu_0 \soneN + \vmu \od  - i \vmu \ox \right) \vn \, , 
\label{simres}
\eea
as a function of \vmu\ and \vn\ as required. 

\section{Example 1: $SU(2)$}
The group $SU(2)$ is used to describe rotations in quantum mechanics and it 
is isomorphic \cite{biedenharn+81,gilmore74a} to the group of unimodular 
quaternions, $Sl(1,q)$. 
The multiplication rules of quaternions being known, explicit expressions for 
the product of two elements of the group $SU(2)$ are obtained easily. In 
quantum mechanics, as a first step one usually establishes the relation 
\bea
\exp [ - i \vA \cdot \vS /2 ] 
   &=& \cos (\alpha/2) \soned - i \sin (\alpha/2) \eA \cdot \vS \, , \\
   & & \vA = \alpha \eA \, , \quad \eA \cdot \eA = 1 \, ,
\label{qmrot}
\eea
by an expansion (\ref{expo}) of the exponential exploiting the simple 
properties of the $(2\times 2)$ Pauli matrices. The three-vector \vA\ 
determines both the axis of rotation, $\eA$, and the turning angle, 
$0 \leq \alpha \leq 4\pi$.  Eq.\ (\ref{qmrot}) is special 
since the matrix in the exponent and the second term on the right are 
proportional to each other. As was mentioned before this is due to the fact 
that the group $SU(2)$ has rank one, implying that all traceless 
$(2 \times 2 )$ matrices are multiples of each other. Working out the product 
of two rotations characterized by \vA\ and \vB , respectively, one obtains   
\bea
\exp [ -i \vG \cdot \vS /2 ] 
  &=& \left( \cos(\alpha/2) \cos(\beta/2) + \vA \cdot \vB \right) \,  
						      \soned  \nonumber \\ 
  & & -i \left( \sin(\alpha/2) \cos(\beta/2) \eA 
       + \cos(\alpha/2) \sin(\beta/2) \eB \right. \nonumber \\
  & &  \left. + \sin(\alpha/2) \sin(\beta/2) \eA \times \eB \right) 
	       \cdot \vS \, .
\label{qmmult}
\eea
The vector \vG\ which points along the axis of the 
composed rotation can be read off directly. 

Eqs. (\ref{qmrot}) and 
(\ref{qmmult}) are derived easily from the spectral method. First,
write down the quantities introduced in the derivation of 
 Eq.\ (\ref{ff}). The spectral theorem (\ref{lst}) involves the projection
operators $\sppm$ (with $(\pm) \equiv (1, 2)$) which for $SU(2)$ are found 
from (\ref{project}) to be
\be
\sppm = \frac{\vA \cdot \vS - \alpha_\mp }{\alpha_\pm - \alpha_\mp}
      = \frac {1}{2} \left( \soned \pm  \eA \cdot \vS \right) \, , 
\label{pro2}      
\ee
using that the operator $\vA \cdot \vS$ has eigenvalues 
$\alpha_\pm = \pm \alpha.$ This immediately reproduces  Eq.\ (\ref{qmrot}) via
\be
e^{-i\alpha_+} \spp +e^{-i\alpha_-} \spm 
       = \exp [ - i \vA \cdot \vS /2 ]     \, .
\label{newrot}
\ee
Writing down the right-hand-side of  Eq.\ (\ref{comp}) for the parameters
$(\mu_0 = \cos(\alpha/2),$ $ \vmu = - \sin(\alpha/2) \eA)$ and similarly for 
$(\nu_0, \vnu)$, one finds that (keep $\od \equiv 0$ in mind)
\bea
\gamma_0 &=&  \cos(\alpha/2) \cos(\beta/2) 
		+\sin(\alpha/2) \sin(\beta/2) \, \eA \cdot \eB \, , \\
\vG      &=&   \left( \cos(\beta/2) \sin(\alpha/2) \, \eA 
	      + \cos(\alpha/2) \sin(\beta/2) \, \eB \right.\nonumber \\
	 & &  \left. + \sin(\alpha/2) \sin(\beta/2) \eA \ox \eB  \right) 
		       \cdot \vS \, .
\label{qmcheck}
\eea
This reproduces indeed Eq. (\ref{qmmult}) because $\ox$ coincides with the 
familiar cross product in three dimensions. Note that the results
have been derived here without explicitly expanding the exponentials involved.

\section{Example 2: $SU(4)$}

The example of $SU(2)$ is exceptional in the sense that 
(i) the product $\od$ is identically zero, (ii) the spectral theorem and its 
linearized version coincide, and (iii) the matrices \sm\ and \ssf\ in  Eq.\  
(\ref{lstip}) are multiples of each other. None of these properties holds 
for $SU(N), N \geq 3$, all of which do provide {\em generic} examples to illustrate 
the BCH-composition rule. Analytic solvability of the third- and fourth-order   
characteristic polynomials is a pleasant accident but it does not have any 
structural consequences in the present context. To give a nontrivial example, 
$SU(4)$ will be studied below.

The interesting point is the reduction of the spectral theorem for an 
element of $SU(4)$ to linear form. Let us assume that the coefficients 
$e_n (\vm)$ of the powers of \sm\ in  Eq.\ (\ref{expo}) have been determined 
(use $f(x) \equiv \exp [-ix]$) by solving the characteristic polynomial of 
\sm\ and employing Eqs. (\ref{highk}) and (\ref{lowk}):
\bea
\exp [ -i \sm ] 
  &=& e_0 \sonef + e_1 \vm \cdot \vsl + e_2 (\vm \cdot \vsl)^2  
					  + e_3 (\vm \cdot \vsl)^3 
		      \label{reduce4}                    \\
  &=& \left( e_1 + e_2 \frac{1}{2} \vm^2 
		 + e_3 \frac{1}{2} (\vm \od \vm ) \cdot \vm \right) \sonef 
		 \nonumber \\
  & &+ \left( (e_1 + e_3 \frac{1}{2} \vm^2 ) \vm + e_2 \vm \od \vm 
		 +  e_3 (\vm \od \vm ) \od \vm \right) \cdot \vsl \, ,
\nonumber
\eea
and that the reduction has been carried out via  Eq.\ (\ref{comcom}), 
using the antisymmetry of the $\ox$ product. Alternatively, one employs 
formula (\ref{ff}) based on the recursion relations. 
The quadratic and cubic terms 
lead to vectors with third powers of \vm\ at most. As an identity, left- and 
right-hand-side of (\ref{reduce4}) must commute which is not trivial only for 
the last two terms multiplying \vsl:
\be
[ \vm \cdot \vsl , (\vm \od \vm ) \cdot \vsl ] 
     = 2i \{\vm \ox (\vm \od \vm ) \} \cdot \vsl = 0 \, ,
\label{three}     
\ee
as follows from (\ref{newjaco2}) applied to the quantity in curly brackets.
Similarly, for the fourth term one finds 
\be
[ \vm \cdot \vsl , \{(\vm \od \vm ) \od \vm \} \cdot \vsl ] 
     = 2i \left( \vm \ox \{(\vm \od \vm ) \od \vm \} \right) \cdot \vsl = 0 \, ,
\label{four}     
\ee
Furthermore, one shows along the same line that these two terms commute 
among themselves,
\bea
[ (\vm \od \vm ) \cdot \vsl , \{(\vm \od \vm ) \od \vm \} \cdot \vsl ] 
     &=& 2i \left( (\vm \od \vm ) \ox 
		\{(\vm \od \vm ) \od \vm \} \right) \cdot \vsl \nonumber \\ 
     &=& 0 \, ,
\label{threefour}     
\eea
Hence, in the process of `linearization,' {\em three} commuting linear 
combinations of the $(N^2 -1)$ matrices \vsl\ arise naturally for $SU(4)$. They span 
the maximal abelian subalgebra associated with the element $\vm \cdot \vsl$. 
Knowing (\ref{reduce4}) it is straightforward to ($i$) multiply two elements 
$\exp [ -i \sm ]$ and $\exp [ -i \sn ]$ of $SU(4)$, ($ii$) reduce the 
product to linear form by removing the single term quadratic in \vsl\ in 
analogy to (\ref{comp}) and to ($iii$) 
reexponentiate using the prescription in (\ref{Rk}).

\section{Summary and Discussion}
It has been shown how to explicitly calculate BCH-relations for the 
group $SU(N)$. The essential ingredients are ($i$) the property that products 
of generators $\slambdaj \in SU(N)$ are expressible as linear combinations of 
generators, and ($ii$) the reduction of the spectral theorem to
linearized form. It has been assumed throughout that the operators involved  
have no degenerate eigenvalues (this case could be included along the 
lines shown in \cite{rhagunathan+92}, for example). The present approach 
is {\em not} restricted to exponential functions of operators which, however, 
seems to be the most important case in physics. Applications of these results 
are expected to deal with coherent states for the group $SU(N)$, useful for 
the description of lasers with $N$ levels. 

Both steps, ($i$) and ($ii$), are based on a surplus of structure in the 
algebra $su(N)$, i. e., the specific form of the anticommutator 
(\ref{anticomm}) which does not exist for all Lie algebras. Therefore, 
the generalization of this approach to other groups is possible whenever 
the {\em product} of two generators defines another element of the original 
algebra. In general, this is guaranteed only for the {\em Lie product}, the 
commutator. To put it differently, the Lie algebra must be closed under both 
commutation {\em and} anticommutation of its elements. Apart from $SU(N)$, 
this property also holds for the general linear group in $N$ dimensions, 
$GL(N)$, for example.

\end{document}